\pdfoutput=1
\RequirePackage{ifpdf}
\ifpdf 
\documentclass[pdftex]{sigma}
\else
\documentclass{sigma}
\fi

\begin{document}
\allowdisplaybreaks

\newcommand{\arXivNumber}{2005.01059}

\renewcommand{\thefootnote}{}

\renewcommand{\PaperNumber}{074}

\FirstPageHeading

\ShortArticleName{The Endless Beta Integrals}

\ArticleName{The Endless Beta Integrals\footnote{This paper is a~contribution to the Special Issue on Elliptic Integrable Systems, Special Functions and Quantum Field Theory. The full collection is available at \href{https://www.emis.de/journals/SIGMA/elliptic-integrable-systems.html}{https://www.emis.de/journals/SIGMA/elliptic-integrable-systems.html}}}

\Author{Gor A.~SARKISSIAN~$^{\dag\ddag\S}$ and Vyacheslav P.~SPIRIDONOV~$^{\dag\S}$}

\AuthorNameForHeading{G.A.~Sarkissian and V.P.~Spiridonov}

\Address{$^\dag$~Laboratory of Theoretical Physics, JINR, Dubna, 141980, Russia}
\EmailD{\href{mailto:sarkissn@theor.jinr.ru}{sarkissn@theor.jinr.ru}, \href{mailto:spiridon@theor.jinr.ru}{spiridon@theor.jinr.ru}}

\Address{$^\ddag$~Department of Physics, Yerevan State University, Yerevan, Armenia}
\EmailD{\href{mailto:gor.sarkissian@ysu.am}{gor.sarkissian@ysu.am}}

\Address{$^\S$~St. Petersburg Department of the Steklov Mathematical Institute of Russian Academy\\
\hphantom{$^\S$}~of Sciences, Fontanka 27, St.~Petersburg, 191023 Russia}

\ArticleDates{Received May 05, 2020, in final form July 24, 2020; Published online August 05, 2020}

\Abstract{We consider a special degeneration limit $\omega_1\to - \omega_2$ (or $b\to {\rm i}$ in the context of $2d$ Liouville quantum field theory) for the most general univariate hyperbolic beta integral. This limit is also applied to the most general hyperbolic analogue of the Euler--Gauss hypergeometric function and its $W(E_7)$ group of symmetry transformations. Resulting functions are identified as hypergeometric functions over the field of complex numbers related to the ${\rm SL}(2,\mathbb{C})$ group. A new similar nontrivial hypergeometric degeneration of the Faddeev modular quantum dilogarithm (or hyperbolic gamma function) is discovered in the limit $\omega_1\to \omega_2$ (or $b\to 1$).}


\Classification{33D60; 33E20}

\renewcommand{\thefootnote}{\arabic{footnote}}
\setcounter{footnote}{0}

\section{Introduction}

The story of beta integrals (the integrals of hypergeometric type admitting exact evaluation)
starts from Euler's proof of the following formula \cite{aar}
\begin{equation}
\int_0^1t^{\alpha-1}(1-t)^{\beta-1}{\rm d}t=\frac{\Gamma(\alpha)\Gamma(\beta)}
{\Gamma(\alpha+\beta)}, \qquad \operatorname{Re}(\alpha), \operatorname{Re}(\beta)>0,
\label{Bfunction}\end{equation}
where $\Gamma(x)$ is the Euler gamma function. Note that the Gaussian integral $\int_{-\infty}^\infty {\rm e}^{-x^2}{\rm d}x=\sqrt{\pi}$ emerges in
a special degeneration limit of this exact relation.
Over the years identity~\eqref{Bfunction} has found many generalizations. In particular, the
$q$-hypergeometric line of developments brought to light the following Askey--Wilson $q$-beta integral
\begin{equation}
\frac{(q;q)_\infty}{4\pi {\rm i}}\int_{{\mathbb T}}
\frac{\big(z^2;q\big)_\infty\big(z^{-2};q\big)_\infty}
{\prod\limits_{j=1}^4(t_jz;q)_\infty\big(t_jz^{-1};q\big)_\infty}
\frac{{\rm d}z}{z}
=\frac{(t_1t_2t_3t_4;q)_\infty}{\prod\limits_{1\le j<k\le 4}(t_jt_k;q)_\infty}, \qquad |q|, |t_j|<1,
\label{AWint}\end{equation}
serving as a measure for the most general classical orthogonal polynomials \cite{aw}.
Here $\mathbb{T}$ denotes the unit circle of counterclockwise orientation
and $(z;q)_\infty:=\prod\limits_{n=0}^\infty\big(1-zq^n\big)$.

Jumping over the Rahman $q$-beta integral \cite{Rahman} extending \eqref{AWint}, we come to the
elliptic beta integral evaluation formula \cite{spi:umn} -- currently the top identity
of the type of interest,
\begin{equation}
\frac{(p;p)_\infty(q;q)_\infty}{4\pi {\rm i}}
\int_\mathbb{T}\frac{\prod\limits_{j=1}^6
\Gamma\big(t_jz^{{\pm 1}};p,q\big)}{\Gamma\big(z^{\pm 2};p,q\big)}\frac{{\rm d}z}{z}
=\prod_{1\leq j<k\leq6}\Gamma(t_jt_k;p,q),
\label{ellbeta}\end{equation}
where $|p|, |q|, |t_j|<1$, $\prod\limits_{j=1}^6t_j=pq$, and
\[
\Gamma(z;p,q):=
\prod_{j,k=0}^\infty\frac{1-z^{-1}p^{j+1}q^{k+1}}{1-zp^{j}q^{k}}
\]
is the elliptic gamma function. Here we apply the standard compact notation
\[
\Gamma\big(tz^{\pm1};p,q\big):=\Gamma(tz;p,q)\Gamma\big(tz^{-1};p,q\big).
\]
The form of this identity is somewhat universal~-- 14 generalized gamma functions
in the integral definition and 15 gamma functions in its exact evaluation expression,
a pattern that will be seen several times below in other instances.

As formally shown in \cite{stok} (see also \cite{DS}), in the limit $|p|, |q| \to 1$ relation \eqref{ellbeta}
reduces to the following hyperbolic beta integral evaluation formula, which is a hyperbolic analogue of
the Rahman integral identity~\cite{Rahman},
\begin{gather}
 \int_{-{\rm i}\infty}^{{\rm i}\infty}
\frac{\prod\limits_{k=1}^6\gamma^{(2)}(g_k\pm z;\mathbf{\omega}) }
{\gamma^{(2)}(\pm 2 z;\mathbf{\omega}) } \frac{{\rm d}z}{2{\rm i}\sqrt{\omega_1\omega_2}}
=
\prod_{1\leq j<k\leq 6}\gamma^{(2)}(g_j+g_k;\mathbf{\omega}),
\label{hyper}\end{gather}
and the following balancing condition holds true
\begin{equation}\label{balcon}
\sum_{k=1}^6 g_k=Q,\qquad Q:=\omega_1+\omega_2.
\end{equation}
Since this identity plays the key role in the following considerations, we shall describe
its ingredients in full detail.

First, we explain the compact notation $\gamma^{(2)}(g\pm u;\mathbf{\omega})
:=\gamma^{(2)}(g+u;\mathbf{\omega})\gamma^{(2)}(g-u;\mathbf{\omega})$, where
\begin{equation}
\gamma^{(2)}(u;\mathbf{\omega})= \gamma^{(2)}(u;\omega_1,\omega_2):={\rm e}^{-\frac{\pi{\rm i}}{2}
B_{2,2}(u;\mathbf{\omega}) } \gamma(u;\mathbf{\omega}),
\label{HGF}\end{equation}
with the second order multiple Bernoulli polynomial
\[
 B_{2,2}(u;\mathbf{\omega})=\frac{1}{\omega_1\omega_2}
\left(\left(u-\frac{\omega_1+\omega_2}{2}\right)^2-\frac{\omega_1^2+\omega_2^2}{12}\right),
\]
and
\begin{equation}
\gamma(u;\mathbf{\omega}):= \frac{\big(\tilde q {\rm e}^{2\pi {\rm i} \frac{u}{\omega_1}};\tilde q\big)_\infty}
{\big({\rm e}^{2\pi {\rm i} \frac{u}{\omega_2}};q\big)_\infty}
=\exp\left(-\int_{{\mathbb R}+{\rm i}0}\frac{{\rm e}^{ux}}
{(1-{\rm e}^{\omega_1 x})(1-{\rm e}^{\omega_2 x})}\frac{{\rm d}x}{x}\right).
\label{int_rep}\end{equation}
The latter function is known as the Faddeev modular quantum dilogarithm \cite{Fad95, Fad94}
or the hyperbolic gamma function~\cite{Ruij}. The relation between its integral and
product forms is described explicitly in~\cite{KLS}, where the inverse of \eqref{int_rep}
was called the double sine function (see also Appendix~A in~\cite{spi:conm} for a
description of different notations used in the literature).

If $\operatorname{Re}(\omega_1), \operatorname{Re}(\omega_2)>0$, then the integral in \eqref{int_rep} converges for $0<\operatorname{Re}(u)< \operatorname{Re}(\omega_1+\omega_2)$. For $\operatorname{Re}(\omega_1), \operatorname{Re}(\omega_2)<0$
it is well defined in the strip $\operatorname{Re}(\omega_1+\omega_2)<\operatorname{Re}(u)< 0$.
For $\operatorname{Re}(\omega_2)\leq 0$ and $\operatorname{Re}(\omega_1)>0$ convergence takes place
for $\operatorname{Re}(\omega_2)<\operatorname{Re}(u)< \operatorname{Re}(\omega_1)$ (by symmetry this is also true
for $\operatorname{Re}(\omega_1)\leq 0$ and $\operatorname{Re}(\omega_2)>0$, if
$\operatorname{Re}(\omega_1)<\operatorname{Re}(u)< \operatorname{Re}(\omega_2)$).
The infinite product representation \eqref{int_rep} is well defined and allows analytical continuation in $u$
to the whole complex plane, provided $|q|<1$, where
\begin{equation}
q= {\rm e}^{2\pi {\rm i}\frac{\omega_1}{\omega_2}},\qquad \tilde q= {\rm e}^{-2\pi {\rm i}\frac{\omega_2}{\omega_1}},
\qquad \text{for} \quad \operatorname{Im}(\omega_1/\omega_2)>0,
\label{q}\end{equation}
or $q= {\rm e}^{2\pi {\rm i}\frac{\omega_2}{\omega_1}}$, $\tilde q= {\rm e}^{-2\pi {\rm i}\frac{\omega_1}{\omega_2}}$, if $\operatorname{Im}(\omega_2/\omega_1)>0$.
Note that the integral representation in \eqref{int_rep} is manifestly symmetric in $\omega_1$ and $\omega_2$
and, moreover, it shows that this function still remains analytical for $\omega_1/\omega_2\in{\mathbb R}/\{0\}$
(i.e., when $|q|=1$) in appropriate domains of~$u$. In the following we stick to the parametrisation \eqref{q}.

Now it is necessary to explain admissible choices of the integration contour in~\eqref{hyper}.
It is not difficult to see that true poles of function $\gamma^{(2)}(u;\mathbf{\omega})$ are located
at the following points
\[
u_p\in \{ -n\omega_1 -m\omega_2\},\qquad n,m \in {\mathbb Z}_{\geq0}.
\]
Therefore poles of the integrand function in \eqref{hyper} form two separate arrays going
to infinity in different directions
\[
z_{\rm poles}\in \{ g_k+n\omega_1+ m\omega_2\}\cup \{ -g_k-n\omega_1- m\omega_2,\},\qquad n,m \in {\mathbb Z}_{\geq0},
\qquad k=1,\ldots, 6.
\]
The contour of integration in \eqref{hyper} should separate these two sets of points.
It remains to explain conditions of the convergence of the integral in~\eqref{hyper}.
For that one should use the following asymptotic formulas~\cite{KLS}:
\begin{gather}\label{asy1}
 A\colon \quad \lim_{z\to \infty}{\rm e}^{{\pi{\rm i}\over 2}B_{2,2}(z;\omega_1,\omega_2)}\gamma^{(2)}(z;\omega_1,\omega_2)=1,
\qquad {\rm for}\quad \operatorname{arg}\omega_1<\operatorname{arg} z<\operatorname{arg}\omega_2+\pi,
\\ \label{asy2}
 B\colon \quad \lim_{z\to \infty}{\rm e}^{-{\pi{\rm i}\over 2}B_{2,2}(z;\omega_1,\omega_2)}\gamma^{(2)}(z;\omega_1,\omega_2)=1,
\qquad {\rm for}\quad \operatorname{arg}\omega_1-\pi<\operatorname{arg} z<\operatorname{arg}\omega_2.
\end{gather}
Applying these formulas to the integrand in \eqref{hyper} when the integration
variable goes to infinity within the indicated cones, one finds the asymptotics
\begin{equation}
A \to {\rm e}^{6\pi{\rm i}\frac{z(\omega_1+\omega_2)}{\omega_1\omega_2}}, \qquad
B \to {\rm e}^{-6\pi{\rm i}\frac{z(\omega_1+\omega_2)}{\omega_1\omega_2}}.
\label{AB}\end{equation}
So, the contour of integration should be chosen in such a way that
both these factors are vanishing exponentially fast.
It is standard to assume that $\operatorname{Re}(\omega_1), \operatorname{Re}(\omega_2)>0$
in which case it is sufficient to take $z\to +{\rm i}\infty$ in the region $A$
and $z\to -{\rm i}\infty$ in the region $B$. Assuming that $\operatorname{Re}(g_k )> 0$,
the contour of integration can be taken as the imaginary axis. After rotating
the integration contour by passing to the integration variable $x= z/\sqrt{\omega_1\omega_2}$
the integral converges for $x\to \pm {\rm i}\infty$, if $\operatorname{Re}(\sqrt{\omega_1/\omega_2})>0$,
and for $\operatorname{Re}(g_k/\sqrt{\omega_1\omega_2})>0$ the imaginary axis of $x$ can be
taken as the integration contour.

For completeness we indicate also the way how formula \eqref{ellbeta} is reduced
to~\eqref{hyper}. Namely, one should parametrise \cite{stok}
\[
t_j={\rm e}^{-2\pi v g_j}, \qquad z={\rm e}^{-2\pi v u},\qquad p={\rm e}^{-2\pi v\omega_1}, \qquad q={\rm e}^{-2\pi v\omega_2}
\]
and take the limit $v\to 0^+$ using the limiting relation
\begin{equation}\label{parlim2}
\Gamma\big({\rm e}^{-2\pi vu};{\rm e}^{-2\pi v\omega_1},{\rm e}^{-2\pi v\omega_2}\big)
\underset{v\to 0^+}{=}
{\rm e}^{-\pi\frac{2u-\omega_1-\omega_2}{12v\omega_1\omega_2}}\gamma^{(2)}(u;\omega_1,\omega_2).
\end{equation}
As shown in~\cite{rai:limits} this transition from the elliptic gamma function to the hyperbolic one is
uniform on compacta. Therefore the degeneration procedure from~\eqref{ellbeta}
to \eqref{hyper} is actually fully legitimate.

We squeezed the history of beta integrals to a few examples and wish to state that it is far from
complete, i.e., its ending is not seen yet, even at the univariate integrals level.
To justify this claim we shall present two more beta integrals which extend the picture in
the directions not expected even after discovery of the elliptic beta integral~\eqref{ellbeta}.

The first new case. Recently we have extended identity \eqref{hyper} to a beta integral associated
with the general lens space~\cite{SS23, SS24}. Corresponding formula has the following form
\begin{gather}
\sum_{m\in {\mathbb Z}_c+\nu}\int_{-{\rm i}\infty}^{{\rm i}\infty}
\frac{\prod\limits_{j=1}^6\Gamma_M(g_j\pm z,n_j\pm m)}
{\Gamma_M(\pm 2z,\pm 2m)} \frac{{\rm d}z}{2{\rm i}c\sqrt{\omega_1\omega_2}}
= \prod_{1\leq \ell<j\leq 6} \Gamma_M(g_\ell+g_j,n_\ell+n_j),
\label{integral}\end{gather}
where ${\mathbb Z}_c = \{0, 1,\ldots,c -1\}$, $n_j\in {\mathbb Z} +\nu$, $\nu = 0,\frac{1}{2}$.
The continuous variables $\omega_{1}$, $\omega_{2}$, $g_j\in {\mathbb C}$,
$\operatorname{Re}(\omega_{1}), \operatorname{Re}(\omega_{2}),\operatorname{Re}(g_j) > 0$,
and discrete ones $n_j$ satisfy the balancing condition
\begin{equation}
\sum_{j=1}^6g_j = \omega_1 + \omega_2,\qquad \sum_{j=1}^6 n_j=-d-1.
\end{equation}
Here $\Gamma_M(\mu\pm z, n\pm m):= \Gamma_M(\mu+z,n+m)\Gamma(\mu-z,n-m)$ and the rarefied hyperbolic gamma function $\Gamma_M(\mu,m)$ has the form
\begin{equation}\label{gammakp}
\Gamma_M(\mu,m):=Z(m){\rm e}^{-\frac{\pi {\rm i}}{2c}B_{2,2}(\mu;\omega_1,\omega_2)}\gamma_M(\mu,m).
\end{equation}
The $\gamma_M(\mu,m)$-function was introduced by Dimofte \cite{Dimofte} as the modular
quantum dilogarithm associated with the general lens space
\begin{equation}
\gamma_M(\mu,m)=\gamma_M(\mu,m;\omega_1,\omega_2):=
\frac{\big(\tilde q {\rm e}^{2\pi {\rm i} \tilde u(\mu, m)};\tilde q\big)_\infty}
{\big({\rm e}^{2\pi {\rm i} u(\mu,m)};q\big)_\infty}, \qquad |q|<1,
\end{equation}
where $q:={\rm e}^{2\pi {\rm i} \tau}$, $\tilde q:={\rm e}^{2\pi {\rm i}\tilde \tau}$,
\[
\tau:=\frac{\omega_1-d\omega_2}{c\omega_2},\qquad \tilde \tau:=\frac{a\tau+b}{c\tau+d}=\frac{a\omega_1-\omega_2}{c\omega_1},
\qquad M=\left(
\begin{matrix}
a & b \\
c & d
\end{matrix} \right)\in {\rm SL}(2,{\mathbb Z}),
\]
and
\[
u(\mu,m):=\frac{\mu+m\omega_2}{c\omega_2}, \qquad \tilde u(\mu,m):=\frac{\mu+am\omega_1}{c\omega_1}
=m\tilde \tau+\frac{u(\mu,m)}{c\tau+d}.
\]
The normalizing factors of \eqref{gammakp} were suggested in \cite{SS23, SS24}. In particular, $Z(m)$ was chosen in
the form
\begin{equation}
Z(m)=\frac{{\rm e}^{-\frac{\pi {\rm i}}{4} (1-\frac{a+d+3}{3c})}}{\varepsilon(a,b,c,d)}
{\rm e}^{\pi {\rm i} \frac{(1+b)c+a}{2c}m(m+d+1)},
\end{equation}
where $\varepsilon(a,b,c,d)$ is a 24-th root of unity emerging in the general
modular transformation law for the Dedekind $\eta$-function
\begin{equation}
\eta\left(\frac{a\tau+b}{c\tau+d}\right)=\varepsilon(a,b,c,d)\sqrt{-{\rm i}(c\tau+d)}\eta(\tau),\qquad
\eta(\tau)={\rm e}^{\frac{\pi {\rm i}\tau}{12}}\big({\rm e}^{2\pi {\rm i}\tau};{\rm e}^{2\pi {\rm i}\tau}\big)_\infty.
\label{eta}\end{equation}
As promised, relation \eqref{integral} contains 14 generalized gamma functions on the
left-hand side and~15 such gamma functions on the right-hand side.

As the second new formula, which is the main goal of this work, we describe a special degeneration limit $\omega_1+\omega_2\to 0$ of the ordinary hyperbolic beta integral~\eqref{hyper}. In a further step, we describe also similar reduction of symmetry transformations of the most general hyperbolic analogue of the Euler--Gauss hypergeometric function.

\section{Gamma function over the complex numbers}

Let us take $\alpha, \alpha'\in\mathbb{C}$ such that $\alpha-\alpha'=n_\alpha \in\mathbb{Z}$
and for $z\in\mathbb{C}$ denote
\[
[z]^\alpha:= z^\alpha \bar z^{\alpha'}=|z|^{2\alpha'} z^{n_\alpha},\qquad
\int_{\mathbb{C}} {\rm d}^2z:=\int_{\mathbb{R}^2}{\rm d}(\operatorname{Re} z)\, {\rm d}(\operatorname{Im} z),
\]
where $\bar z$ is a complex conjugate of $z$.
Then one has the following complex beta integral evaluation formula \cite{GGV}
\begin{equation}\label{CB}
\int_{\mathbb{C}}[w-z_1]^{\alpha-1} [z_2-w]^{\beta-1} \frac{{\rm d}^2w}{\pi}
=\frac{\Gamma(\alpha)\Gamma(\beta)}{\Gamma(\alpha+\beta)}
\frac{\Gamma(1-\alpha'-\beta')}{\Gamma(1-\alpha')\Gamma(1-\beta')}[z_2-z_1]^{\alpha+\beta-1},
\end{equation}
which is a clear analogue of relation~\eqref{Bfunction}.

This formula suggests the definition of a gamma function over the field of complex numbers as
a particular ratio of Euler's gamma functions
\begin{equation}
{\bf \Gamma}(x,n)={\bf \Gamma}(\alpha|\alpha'):=\frac{\Gamma(\alpha)}{\Gamma(1-\alpha')}
=\frac{\Gamma\big(\frac{n+{\rm i}x}{2}\big)}{\Gamma\big(1+\frac{n-{\rm i}x}{2}\big)},
\qquad \alpha=\frac{n+{\rm i}x}{2},\qquad \alpha'=\frac{-n+{\rm i} x}{2},
\label{Cgamma}\end{equation}
where $x\in {\mathbb C}$ and $n\in {\mathbb Z}$.
From the reflection relation for the Euler gamma function $\Gamma(x)\Gamma(1-x)=\pi/\sin\pi x$,
it follows that
\begin{equation}
{\bf \Gamma}(\alpha|\alpha') =(-1)^{\alpha-\alpha'}{\bf \Gamma}(\alpha'|\alpha), \qquad
{\bf \Gamma}(x,-n)=(-1)^n{\bf \Gamma}(x,n),
\end{equation}
and
\begin{equation}
{\bf \Gamma}(\alpha|\alpha'){\bf \Gamma}(1-\alpha|1-\alpha') =(-1)^{\alpha-\alpha'}, \qquad
{\bf \Gamma}(x,n){\bf \Gamma}(-x-2{\rm i},n)=1.
\label{reflCgamma}\end{equation}
The functional equation takes the form
\[
{\bf \Gamma}(\alpha+1|\alpha'+1) =-\alpha\alpha' {\bf \Gamma}(\alpha|\alpha'),\qquad
{\bf \Gamma}(x-2{\rm i},n)=\frac{n^2+x^2}{4}{\bf \Gamma}(x,n).
\]

Now one can rewrite the right-hand side of \eqref{CB} in the following forms
\[
\frac{{\bf\Gamma}(\alpha|\alpha'){\bf\Gamma}(\beta|\beta')} {{\bf\Gamma}(\alpha+\beta|\alpha'+\beta')}
[z_2-z_1]^{\alpha+\beta-1}= \frac{{\bf\Gamma}(\alpha,\beta,\gamma)}{[z_1-z_2]^{\gamma}},
\qquad {\bf\Gamma}(\alpha_1,\ldots,\alpha_k):=\prod_{j=1}^k{\bf\Gamma}(\alpha_j|\alpha_j'),
\]
where $\alpha+\beta+\gamma= \alpha'+\beta'+\gamma'=1$.

After making the inversion transformations $w\to w^{-1}$, $z_1\to z_1^{-1}$, $z_2 \to z_2^{-1}$
and the shifts $w\to w-z_3$, $z_1\to z_1-z_3$, $z_2\to z_2-z_3$, relation \eqref{CB}
takes the form of a star-triangle relation:
\begin{gather}
\int_{\mathbb{C}}[z_1-w]^{\alpha-1} [z_2-w]^{\beta-1} [z_3-w]^{\gamma-1}\frac{{\rm d}^2w}{\pi}
\nonumber\\
\qquad{} =\frac{ {\bf\Gamma}(\alpha,\beta,\gamma) } {[z_3-z_2]^{\alpha}[z_1-z_3]^{\beta}[z_2-z_1]^{\gamma}},
\qquad \alpha+\beta+\gamma=1. \label{STR}
\end{gather}
Multidimensional analogues of complex integrals \eqref{CB} and \eqref{STR} were considered
by Dotsenko and Fateev within the context of $2d$ conformal field theory~\cite{DF}. An independent study of the
complex Selberg integral was performed in~\cite{Aomoto}. Such integrals naturally emerge also in the
theory of non-compact ${\rm SL}(2, {\mathbb C})$ spin chains~\cite{DMV2017,DMV2018}.

The well known trigonometric $q$-gamma function \cite{aar} is defined as
\begin{equation}
\Gamma_q(x):=\frac{(q;q)_\infty}{(q^x;q)_\infty}(1-q)^{1-x},\qquad |q|<1,\qquad x\in\mathbb{C}.
\end{equation}
For fixed $x$, in the limit $q\to 1^-$ one obtains the Euler gamma function
\begin{equation}\label{gamlim}
\lim_{q\to 1^-} \Gamma_q(x)= \Gamma(x).
\end{equation}
As shown in \cite{Tom,rai:limits}, this convergence to gamma function is uniform on compacta,
which allows degeneration of the $q$-beta integral \eqref{AWint} with a compact measure support
to the de Branges--Wilson integral \cite{aar} with an infinite Mellin--Barnes type integration contour.

We would like to consider now a similar limit for the hyperbolic gamma function \eqref{int_rep}
\begin{equation}
\gamma(u;\omega_1,\omega_2)={\big({\rm e}^{2\pi {\rm i}\frac{u}{\omega_1}}{\rm e}^{-2\pi {\rm i} {\omega_2\over \omega_1}};{\rm e}^{-2\pi {\rm i} {\omega_2\over \omega_1}}\big)_{\infty}\over
\big({\rm e}^{2\pi {\rm i} \frac{u}{\omega_2}};{\rm e}^{2\pi {\rm i} {\omega_1\over \omega_2}}\big)_{\infty}}.
\label{hypergamma}\end{equation}

In the context of $2d$ quantum Liouville theory it is customary to use notations \cite{DF,ILT,RW}
\[
b:=\sqrt{\omega_1\over \omega_2}, \qquad q={\rm e}^{2\pi {\rm i} b^2},\qquad \tilde q={\rm e}^{-2\pi {\rm i} b^{-2}}.
\]
The central charge $c$ of this theory, the key characteristic of $2d$ conformal field theory \cite{DF},
has the form $c=1+6\big(b+b^{-1}\big)^2$, i.e., $b$ is the variable parametrising~$c$.
Let us consider the cases when simultaneously $q\to 1$ and $\tilde q\to 1$,
so that in~\eqref{hypergamma} there starts to emerge a ratio of the Euler gamma functions.
Clearly this is possible, if $b^2\to n$, $b^{-2}\to m$, $n, m\in {\mathbb Z}$.
Evidently the only admissible choices are $b=\pm{\rm i}$, when $c=1$
(this case can be considered as a $p\to\infty$ limit of the minimal models~\cite{RW}), and
$b=\pm1$, when $c=25$.

Consider the first of these possibilities (the case $b=\pm 1$ will be considered in Section~\ref{newlimit}).
Namely, let us take small $\delta>0$ and set
\begin{equation}\label{om1om22}
b=\sqrt{\omega_1\over \omega_2}={\rm i}+\delta, \qquad \delta\to 0^+.
\end{equation}
Obviously one has now
\begin{equation}\label{om1om2}
\sqrt{\omega_2\over \omega_1}=-{\rm i}+\delta+O\big(\delta^2\big),\qquad
{\omega_1\over \omega_2}=-1+2{\rm i}\delta+\delta^2,\qquad {\omega_2\over \omega_1}=-1-2{\rm i}\delta+O\big(\delta^2\big),
\end{equation}
as well as $Q=\omega_1+\omega_2=2\delta\sqrt{\omega_1\omega_2}+O\big(\delta^2\big).$
In addition to this choice, we parametrise the argument $u$ in \eqref{hypergamma} as follows
\begin{equation}
u={\rm i}\sqrt{\omega_1\omega_2}(n+x\delta), \qquad n\in {\mathbb Z}, \qquad x\in {\mathbb C},
\end{equation}
and consider the limit $\delta\to 0^+$. Let us investigate behavior of each of the
infinite products in~\eqref{hypergamma}. In the denominator we have
\[
\big({\rm e}^{2\pi {\rm i} \frac{u}{\omega_2}};{\rm e}^{2\pi {\rm i} {\omega_1\over \omega_2}}\big)_{\infty}
=\big({\rm e}^{-2\pi\delta(n+{\rm i}x+\delta x)};q\big)_\infty
=\frac{(q;q)_\infty(1-q)^{1-\frac{n+{\rm i}x}{2}+O(\log q)}}
{\Gamma_q\big(\frac{n+{\rm i}x}{2}+O(\log q)\big)},
\]
where $q={\rm e}^{-4\pi\delta(1-{\rm i}\delta/2)}$. Analogously, for the numerator we obtain
\[
\big({\rm e}^{2\pi {\rm i}\frac{u}{\omega_1}}{\rm e}^{-2\pi {\rm i} {\omega_2\over \omega_1}};{\rm e}^{-2\pi {\rm i} {\omega_2\over \omega_1}}\big)_{\infty}
= \big(\tilde q^{1+\frac{n-{\rm i}x}{2}+O(\log \tilde q)};\tilde q\big)_\infty
=\frac{(\tilde q;\tilde q)_\infty(1-\tilde q)^{\frac{-n+{\rm i}x}{2}+O(\log \tilde q)}}
{\Gamma_{\tilde q}\big(1+\frac{n-{\rm i}x}{2}+O(\log \tilde q)\big)},
\]
where
\[
\tilde q={\rm e}^{-4\pi\delta\frac{1-{\rm i}\delta/2}{(1-{\rm i}\delta)^2}}.
\]
As a result we obtain,
\[
\gamma(u;\omega_1,\omega_2)=\frac{\Gamma_q\big( \frac{n+{\rm i}x}{2}+O(\log q) \big)}
{\Gamma_{\tilde q}\left( 1+\frac{n-{\rm i}x}{2}+O(\log \tilde q)\right)}
\frac{(\tilde q;\tilde q)_\infty}{(q;q)_\infty}
\frac{(1-\tilde q)^{\frac{-n+{\rm i}x}{2}+O(\log \tilde q)}}
{(1-q)^{1-\frac{n+{\rm i}x}{2}+O(\log q)}}.
\]

Now we apply a slightly stronger limiting relation than \eqref{gamlim}, namely
for $q\to 1$ we take $\Gamma_q(x+O(\log q))\to \Gamma(x)$.
As follows from the considerations of~\cite{Tom} (see there Appendix~B) and \cite{rai:limits}
this limit is uniform over the compact domains excluding poles similar to \eqref{gamlim}
for real $q\to 1^-$ and complex $x$ away from the poles. The arguments of \cite{rai:limits}
show that this property is preserved even for complex~$q$, provided it approaches~1
with the angle away from $\pm \pi/2$, which is satisfied in our case, since this angle
is proportional to~$\delta$. This uniformity
will be very useful for consideration of such a limit for hyperbolic integrals.

Using the modular transformation rule $\eta(-1/\tau)=\sqrt{-{\rm i}\tau} \eta(\tau)$
for the Dedekind eta-func\-tion~\eqref{eta}, we find
\begin{equation}
\frac{(\tilde q;\tilde q)_\infty}{(q;q)_\infty}
={\rm e}^{{\pi{\rm i} \over 12}\big({\omega_2\over \omega_1}+{\omega_1\over \omega_2}\big)}
\left(-{\rm i} {\omega_1\over \omega_2}\right)^{1\over 2}
\underset{\delta\to 0^+}{=}{\rm e}^{ \pi{\rm i} \over 12}.
\label{ratio}\end{equation}
Note that for $\delta\to 0^+$ one has $\tilde q= q+O\big(\delta^2\big)$, but formal substitution of
this relation to $(\tilde q;\tilde q)_\infty$ in \eqref{ratio} and
termwise cancellation of the individual multipliers in the ratio of infinite products
of interest yields 1, instead of the nontrivial phase factor ${\rm e}^{ \pi{\rm i} \over 12}$.

Finally, we come to the leading asymptotics
\begin{equation}
\gamma(u;\omega_1,\omega_2)\underset{\delta\to 0^+}{=} {\rm e}^{\pi {\rm i}\over 12}
(4\pi\delta)^{{\rm i}x-1}{\bf \Gamma}(x,n),
\end{equation}
where ${\bf \Gamma}(x,n)$ is the complex gamma function defined in~\eqref{Cgamma}.
This result was presented first in~\cite{BMS} without any derivation details and
rigorous justifications. Recalling definition (\ref{HGF}), we obtain
\begin{equation}\label{gam2lim2}
\gamma^{(2)}({\rm i}\sqrt{\omega_1\omega_2}(n+x\delta);\omega_1,\omega_2)\underset{\delta\to 0^+}{=} {\rm e}^{\frac{\pi {\rm i}}{2}n^2} (4\pi\delta)^{{\rm i}x-1}{\bf \Gamma}(x,n),
\qquad \sqrt{\omega_1\over \omega_2}={\rm i}+\delta,
\end{equation}
where $n\in {\mathbb Z}$, $x\in {\mathbb C}$.
Thus in this limit the function $\gamma^{(2)}(u;\mathbf{\omega})$
starts to blow up around a~special discrete set of points of the argument $u$
passing through the whole complex plane along a particular line.

Note that the choice $b= -{\rm i}$ is equivalent to \eqref{om1om22}:
the ansatz $b:=-{\rm i}+\delta$,
$\delta<0$, yields for $\delta\to 0^-$ the same limit as in~\eqref{gam2lim2}
and leads to results identical to the ones described below for~\eqref{om1om22}.

\section[General complex beta integral ($\omega_1=-\omega_2$)]{General complex beta integral ($\boldsymbol{\omega_1=-\omega_2}$)}\label{genbet}

We are going to apply the limit considered in the previous section to an integral of the form
\begin{equation}\label{intdel}
\int_{-{\rm i}\infty}^{{\rm i}\infty}\Delta(z){{\rm d}z\over {\rm i}\sqrt{\omega_1\omega_2}}
=\int_{-{\rm i}\infty}^{{\rm i}\infty}\Delta(\sqrt{\omega_1\omega_2}x){{\rm d}x\over {\rm i}},
\qquad x=\frac{z}{\sqrt{\omega_1\omega_2}},
\end{equation}
where $\Delta(z)$ is a product of $\gamma^{(2)}(u;\omega_1,\omega_2)$ functions.
Here we assume that these integrals converge when the integration contour is taken
as the imaginary axis for both integration variables~$z$ and~$x$, i.e., the integration
contour can be rotated by the angle $\arg \sqrt{\omega_1\omega_2}$.
The function $\gamma^{(2)}(u;\omega)$ is uniform, i.e., we can scale all variables and it
does not change the form of this function,
\[
\gamma^{(2)}(\lambda u;\lambda\omega_1,\lambda\omega_2)=\gamma^{(2)}(u;\omega_1,\omega_2), \qquad \lambda\neq 0,
\]
Therefore we can fix the product $\omega_1\omega_2$ to be any nonzero number.
E.g., in the quantum Liouville theory the standard normalization is $\omega_1\omega_2=1$, or $\omega_1=b$, $\omega_2=b^{-1}$
\cite{ILT,RW, DS}, which corresponds to the choice $\lambda=1/\sqrt{\omega_1\omega_2}$.

First, we rewrite this integral as an infinite sum
\begin{gather*}
\int_{-{\rm i}\infty}^{{\rm i}\infty}\Delta\big(\sqrt{\omega_1\omega_2} x\big){{\rm d}x\over {\rm i}}=
\sum_{N\in {\mathbb Z}} \int_{{\rm i}(N-1/2)}^{{\rm i}(N+1/2)}\Delta\big(\sqrt{\omega_1\omega_2} x\big){{\rm d}x\over {\rm i}}\\
\hphantom{\int_{-{\rm i}\infty}^{{\rm i}\infty}\Delta\big(\sqrt{\omega_1\omega_2} x\big){{\rm d}x\over {\rm i}}}{}
= \sum_{N\in {\mathbb Z}} \int_{N-1/2}^{N+1/2}\Delta\big( {\rm i}\sqrt{\omega_1\omega_2} x\big){\rm d}x
=\sum_{N\in {\mathbb Z}} \int_{-1/2}^{1/2} \Delta\big({\rm i}\sqrt{\omega_1\omega_2}(N+x)\big){\rm d}x.
 \end{gather*}
In the last two steps we changed the variable $x\to {\rm i} x$
with the subsequent shift $x\to x+N$.
Now we parametrise $x=y\delta$, $\delta>0,$ and take the limit $\delta\to 0^+$.
Then we have
\[
 \lim_{\delta\to 0}\sum_{N\in {\mathbb Z}} \int_{-1/2}^{1/2}
\Delta\big({\rm i}\sqrt{\omega_1\omega_2}(N+x)\big){\rm d}x= \lim_{\delta\to 0}
\sum_{N\in {\mathbb Z}} \int_{-1/2\delta}^{1/2\delta}
\delta \Delta\big({\rm i}\sqrt{\omega_1\omega_2}(N+y\delta)\big){\rm d}y.
\]
The sum over $N$ is infinite and for $\delta\to 0^+$ the integration contour becomes the
noncompact real axis $(-\infty,\infty)$. Therefore, in order to interchange
the $\lim\limits_{\delta\to 0}$ sign with the summation and integration, we need the uniform
convergence of the limit \eqref{gamlim}, which is true and thus justifies our
formal manipulations. Finally, we obtain
\begin{equation}
\sum_{N\in {\mathbb Z}}
\int_{-\infty}^{\infty} \Big[
 \lim_{\delta\to 0}\delta\Delta\big({\rm i}\sqrt{\omega_1\omega_2}(N+y\delta)\big)\Big] {\rm d}y.
\label{integral_result}\end{equation}
So, the behaviour of the initial integral (\ref{intdel}) for $\delta\to 0$ is determined
by the asymptotics of integrands in \eqref{integral_result}, provided they are well defined.

Now we apply this reasoning to the beta integral \eqref{hyper}.
Besides taking the integration variable in the above mentioned form
\begin{equation}\label{zkom}
z= {\rm i}\sqrt{\omega_1\omega_2}(N+\delta y),
\qquad y\in {\mathbb C},\qquad N\in {\mathbb Z}+\nu,\qquad \nu=0, \frac{1}{2},
\end{equation}
where $\delta$ is taken to $0^+$,
we scale also the parameters $g_k$ according to the same rule,
\begin{equation}\label{gkom}
g_k={\rm i}\sqrt{\omega_1\omega_2}(N_k+\delta a_k),\qquad a_k\in {\mathbb C},
\qquad N_k\in {\mathbb Z}+\nu,\qquad \nu=0, \frac{1}{2},
\end{equation}
where $a_k$ and $N_k$ satisfy the constraints
\begin{equation}
\sum_{k=1}^6 a_k=-2{\rm i},\qquad \sum_{k=1}^6 N_k=0,
\end{equation}
 following from the balancing condition~\eqref{balcon}.
Note the appearance of a new discrete parameter $\nu=0, \frac{1}{2}$ in formulas \eqref{zkom}
and~\eqref{gkom}. It emerges from the
fact that only the sums $N+N_k$ or the differences $N-N_k$ should be integers in the
arguments of the hyperbolic gamma functions, as required in the limit \eqref{gam2lim2}.
Considerations given in the derivation of formula \eqref{integral_result} remain valid
after the replacement of $N$ by $N+\nu$, $\nu=0, \frac{1}{2}$, i.e., we can replace summations
over $N\in {\mathbb Z}$ by the sums over $N\in {\mathbb Z}+\nu$.

In terms of the integration variable $x=z/\sqrt{\omega_1\omega_2}$ the
original integrand asymptotics \eqref{AB} for $x\to \pm {\rm i}\infty$ takes the form
${\rm e}^{- 12\pi \delta |x|}$, i.e., for finite $\delta$ the integral does converge.
However, for $\delta\to 0^+$ is starts to diverge and we need to estimate
the rate of this divergence.

Inserting parametrisations (\ref{zkom}) and (\ref{gkom}) in (\ref{hyper}), and recalling
 the asymptotics (\ref{gam2lim2}), we find the limiting relations
\begin{gather}
\prod_{k=1}^6\gamma^{(2)}(g_k\pm z;\mathbf{\omega})\to \frac{(-1)^{2\nu}}{(4\pi\delta)^{8}}
\prod_{k=1}^6{\bf \Gamma}(a_k+y,N_k+N){\bf \Gamma}(a_k-y,N_k-N),\nonumber\\
\prod_{1\leq j < k\leq 6}
\gamma^{(2)}(g_j+g_k;\mathbf{\omega})\to \frac{(-1)^{2\nu}}{(4\pi\delta)^{5}}\prod_{1\leq j < k \leq 6}
{\bf \Gamma}( a_j+a_k,N_j+N_k),\nonumber\\
\gamma^{(2)}(\pm 2z;\mathbf{\omega}) \to \frac{(-1)^{2\nu}}{(4\pi\delta)^{2}}
{\Gamma(N+ {\rm i}y)\over \Gamma(1+N- {\rm i}y)}
{\Gamma(-N- {\rm i}y)\over \Gamma(1-N+ {\rm i}y)}={(4\pi\delta)^{-2}\over y^2+N^2}.
\label{extrasign}
\end{gather}
Collecting all the multipliers and cancelling the diverging factor $(4\pi\delta)^{-5}$ on both
sides of the equality \eqref{hyper}, we obtain our key complex beta integral:
\begin{equation}
\frac{1}{8\pi}\sum_{N\in {\mathbb Z}+\nu}\int_{-\infty}^{\infty}\big(y^2+N^2\big)
\prod_{k=1}^6{\bf \Gamma}(a_k\pm y,N_k\pm N)dy=\prod_{1\leq j< k \leq 6}{\bf \Gamma}( a_j+a_k,N_j+N_k),
\label{keybeta}\end{equation}
where $\sum\limits_{k=1}^6 a_k=-2{\rm i}$, $\sum\limits_{k=1}^6 N_k=0$, and
\[
{\bf \Gamma}(x_1\pm x_2 ,n_1\pm n_2):={\bf \Gamma}(x_1+ x_2 ,n_1+ n_2){\bf \Gamma}(x_1- x_2 ,n_1- n_2) .
\]
Here we have the variables $N_k$, $N \in {\mathbb Z}+\nu$, $\nu=0, \frac{1}{2}$, so that
their sums or differences take integer values. Again, this formula contains 14 generalized gamma
functions on the left-hand side and~15 of them on the right-hand side. It can be considered as
a complex analogue of the plain hypergeometric Rahman beta integral~\cite{Rahman}, since it also
contains five free continuous (complex) parameters (five discrete parameters are associated with
them in accordance to the ${\rm SL}(2,{\mathbb C})$ principal series representations parametrization).

Formal poles of the integrands in \eqref{keybeta} are located at the points
\[
y^{(1)}_p\in\{{\rm i}(N+N_k)-a_k+2{\rm i}\ell_1\},
\qquad y^{(2)}_p\in \{-{\rm i}(N_k-N)+a_k-2{\rm i}\ell_2\},
\qquad \ell_1, \ell_2\in {\mathbb Z}_{\geq 0},
\]
and the corresponding formal zeros are
\begin{gather*}
y^{(1)}_z\in \{-2{\rm i}-{\rm i}(N_k+N)-a_k-2{\rm i}\ell_3\},\\
 y^{(2)}_z \in \{2{\rm i}+{\rm i}(N_k-N)+a_k+2{\rm i}\ell_4\},
\qquad \ell_3, \ell_4\in {\mathbb Z}_{\geq 0}.
\end{gather*}
The sets $y^{(1)}_p$ and $y^{(2)}_z$ (or $y^{(2)}_p$ and $y^{(1)}_z$) may overlap
only if simultaneously $\operatorname{Re}(a_k)=0$ and $\operatorname{Im}(a_k)\in {\mathbb Z}$. Let us demand that
$\operatorname{Im}(a_k)\notin {\mathbb Z}/\{0\}$ and discuss the special case of real $a_k$ separately.
Analyzing the overlap of $y^{(1)}_p$ with $y^{(1)}_z$ jointly with the overlap
of $y^{(2)}_p$ with $y^{(2)}_z$, leading to cancellations of poles and zeros,
we come to the conclusion that for any~$N$ true poles on the integrand are located at
\begin{equation}
y_{\rm poles} \in \{ {\rm i}|N+N_k|-a_k+2{\rm i}\ell_1,
-{\rm i}|N-N_k|+a_k-2{\rm i}\ell_2\}, \qquad \ell_1, \ell_2\in {\mathbb Z}_{\geq 0}.
\end{equation}

Therefore for the choice $\operatorname{Im}(a_k)<0$ (which follows from the conditions
$\operatorname{Re}\big(g_k/\sqrt{\omega_1\omega_2}\big)>0$ and $\delta\to 0^+$) the real axis separates sequences of
 poles going to infinity upwards from the ones falling down and the derived formula is true under
these conditions. As to the case $a_k\in {\mathbb R}$, we can perform
analytical continuation. Namely, we deform the contour of integration slightly below the
real axis in such a way that for $\operatorname{Im}(a_k)=0$ no poles emerge on the integration contour
and the formula remains true in this case as well.

For $\nu=0$ the functions of the type standing in the left-hand side of~\eqref{keybeta} appeared
for the first time in
Naimark's investigation of the representation theory of the Lorentz group ${\rm SL}(2,{\mathbb C})$~\cite{Naimark}.
In particular, for $a_k\in {\mathbb R}$ one deals with the unitary principal series representation of this group.
In the modern time, continuation of the investigation of such functions has been launched
by Ismagilov~\cite{Ismag2}, who constructed $6j$-symbols for the ${\rm SL}(2, {\mathbb C})$ group
(for a~verification of his result reached via a different approach, see~\cite{DS2017}). As shown
in~\cite{DMV2017} (see there Appendix~B),
a particular subcase of relation \eqref{keybeta} (see details below) corresponds to the Mellin--Barnes
representation of the complex beta integral \eqref{STR}, which is equivalent to identity \eqref{CB} initially
considered in~\cite{GGV}. The first understanding that such mathematical structures emerge as a special
limit of $q$-hypergeometric functions defined with the help of Faddeev's
 modular quantum dilogarithm, or the hyperbolic gamma function, was reached in~\cite{BMS}.
Thus, joint efforts of the works \cite{Ismag2} and \cite{BMS,DMV2017} have shown that the representation
theory of Faddeev's modular double \cite{fad:mod} comprises the representation theory of the ${\rm SL}(2,{\mathbb C})$
group. A rigorous consideration of the Hilbert space aspects of this class of special functions of
hypergeometric type is given in \cite{MN,Neretin2019}. Another recent related study can be found in~\cite{Mimachi}.

The class of functions emerging for $\nu=1/2$ is a new one and its group-theoretical understanding
is still missing. For the first time existence of such a nontrivial additional discrete parameter was noticed
in \cite{spi:rare} in the investigation of elliptic hypergeometric functions related to the lens space
(such functions were considered also in \cite{KY}),
where the choice $\nu=1/2$ resulted in the discovery of a novel family of trigonometric
$q$-hypergeometric integrals.
A similar situation holds true for the rarefied hyperbolic functions described
above \eqref{integral} \cite{SS23, SS24}. Existence of the discrete variable $\nu=1/2$
in the Mellin--Barnes type representation of complex hypergeometric integrals was
noticed first in \cite{DMV2018}.

As shown in \cite{spi:conm}, the original hyperbolic beta integral evaluation formula \eqref{hyper}
can be represented in the star-triangle form useful for solving the Yang--Baxter equation \cite{CS}.
Therefore its limiting relation we have derived \eqref{keybeta} also can be written in this attractive
form which is useful for solvable models in statistical mechanics.
A special case of identity \eqref{hyper} corresponding to $\nu=0$ and a reduced number of
discrete parameters $N_k$ appeared first in \cite{Kels2014} exactly in the form of the star-triangle
relation.\footnote{After presenting relation \eqref{keybeta} at the Nordita conference in June 2019, there appeared the work by Derkachov and Manashov \cite{DM2019} where it was independently derived (as well as its substantially
more complicated multidimensional version) by a completely different method.
Namely, our identity \eqref{keybeta} corresponds to the choice $N=2$ in formulas (3.7) and (3.8)
in~\cite{DM2019} (formula~(3.8) was obtained from~(3.7) after applying the reflection formula
\eqref{reflCgamma} to the $\bf \Gamma$-functions depending on $x_6$ with a small typo
on the right-hand side, where $2N+3$ should be replaced by $2N+1$). The difference in the sign
factors on the right-hand sides emerges from different representations of the product of complex
gamma functions in the kernel denominator \eqref{extrasign}. Since the star-triangle relation form
of~\eqref{keybeta} was considered in detail in~\cite{DM2019}, we skip its discussion here.}

Beta integrals play a key role in the construction of symmetry transformations for higher order hypergeometric
functions of the corresponding type. At the top elliptic level such consequences of identity \eqref{ellbeta}
were considered in \cite{spi:theta} and in a more general setting in \cite{spi:rare,spi:rareYBE}.
Let us derive symmetry transformations for the top complex hypergeometric function
generalizing the Euler--Gauss $_2F_1$-function by reducing such transformations for
an elliptic hypergeometric function.

\section{Transformation rule I}

Consider the $V$-function, an elliptic analogue of the Euler--Gauss hypergeometric function \cite{spi:essays},
\begin{equation}
V(t_1,\ldots,t_8;p,q)=\frac{(p;p)_\infty(q;q)_\infty}{4\pi {\rm i}}\int_{\mathbb{T}}
{\prod\limits_{a=1}^8\Gamma\big(t_az^{\pm1}; p,q\big)\over \Gamma\big(z^{\pm 2}; p,q\big)}{{\rm d}z\over z},
\end{equation}
where the parameters satisfy constraints $|t_a|<1$ and the balancing condition
$ \prod\limits_{a=1}^8t_a=p^2q^2$.
This function has the $W(E_7)$ Weyl group symmetry transformations, whose key generating
relation has been established in \cite{spi:theta}:
\begin{gather}\label{vt1}
V(t_1,\ldots,t_8;p,q)=\prod_{1\leq j< k\leq 4}\Gamma(t_jt_k;p,q)
\prod_{5\leq j< k\leq 8}\Gamma(t_jt_k;p,q) V(s_1,\ldots,s_8;p,q),
\end{gather}
where
\begin{equation}
s_j=\rho^{-1}t_j,\qquad s_{j+4}=\rho t_{j+4},\qquad j=1,2,3,4, \qquad
\rho=\sqrt{t_1t_2t_3t_4\over pq}=\sqrt{pq\over t_5t_6t_7t_8}.
\end{equation}

Consider the function $I_h(\underline{g})$ defined by the integral
\begin{equation}
I_h(\underline{g})=\int_{-{\rm i}\infty}^{{\rm i}\infty}{\prod\limits_{j=1}^8\gamma^{(2)}(g_j\pm z;\omega_1,\omega_2)\over
\gamma^{(2)}(\pm 2z;\omega_1,\omega_2)}\frac{{\rm d}z}{2{\rm i}\sqrt{\omega_1\omega_2}},
\end{equation}
with $g_j$ satisfying the conditions $\operatorname{Re}(g_j)>0$ and
\begin{equation}\label{mu8}
\sum_{j=1}^8 g_j=2Q,\qquad Q:=\omega_1+\omega_2.
\end{equation}
This is the most general hyperbolic analogue of the Euler--Gauss hypergeometric $_2F_1$-function
satisfying a second order difference equation. It represents a one-parameter extension of the
function built in \cite{Ruij}.

Applying the hyperbolic degeneration limit \eqref{parlim2} to the transformation rule \eqref{vt1},
one comes to the following relation \cite{BRS}
\begin{equation}\label{ide1}
I_h(\underline{g})=\prod_{1\leq j< k \leq 4}\gamma^{(2)}(g_j+g_k;\omega_1,\omega_2)
\prod_{5\leq j< k \leq 8}\gamma^{(2)}(g_j+g_k;\omega_1,\omega_2) I_h(\underline{\lambda}) ,
\end{equation}
where
\begin{equation}
\lambda_j=g_j+\xi, \qquad \lambda_{j+4}=g_{j+4}-\xi, \qquad j=1,2,3,4,\qquad
\xi={1\over 2}\left(\omega_1+\omega_2-\sum_{j=1}^4 g_j\right).
\end{equation}

As a next step, let us take the parametrisations \eqref{om1om22},
\eqref{zkom} and \eqref{gkom} of variables in (\ref{ide1}) and consider the limit $\delta\to 0^+$.
The balancing condition (\ref{mu8}) passes to the following constraints
\begin{equation}\label{balanca}
\sum_{k=1}^8 a_k=-4{\rm i},\qquad a_k\in {\mathbb C},\qquad \sum_{k=1}^8 N_k=0,
\qquad N_k\in {\mathbb Z}+\nu, \qquad \nu=0,\frac{1}{2}.
\end{equation}
This leads to the parametrisation
\begin{equation}
{\xi\over {\rm i}\sqrt{\omega_1\omega_2}}=-{L\over 2}+\delta\left(-{\rm i}-{X\over 2}\right),\qquad
X:=\sum_{j=1}^4 a_j, \qquad L:=\sum_{j=1}^4 N_j.
\end{equation}
Using considerations of the previous section, now it is straightforward to see
that the asymptotic formula (\ref{gam2lim2}) allows the $\delta\to 0^+$ reduction
of identity (\ref{ide1}) to the following symmetry transformation relation
\begin{gather} \nonumber
\sum_{N\in {\mathbb Z}+\nu}\int_{-\infty}^{\infty}\left(y^2+N^2\right) \prod_{k=1}^8{\bf \Gamma}(a_k\pm y,N_k\pm N){\rm d}y
\\ \nonumber
\qquad{} =(-1)^L\prod_{1\leq j< k \leq 4}{\bf \Gamma}(a_j+a_k,N_j+N_k)\prod_{5\leq j< k \leq 8}{\bf \Gamma}( a_j+a_k,N_j+N_k)
\\ \nonumber \qquad\quad{} \times
\sum_{N\in {\mathbb Z}+\mu}\int_{-\infty}^{\infty}\big(y^2+N^2\big)
\prod_{k=1}^4{\bf \Gamma}\big(a_k\pm y- \tfrac{1}{2}X-{\rm i},N_k\pm N-\tfrac{1}{2}L\big)
\\ {} \qquad\quad{}\times
\prod_{k=5}^8{\bf \Gamma}\big(a_k\pm y+\tfrac{1}{2}X+{\rm i},N_k\pm N+\tfrac{1}{2}L\big){\rm d}y,
\label{complextrafoI}
\end{gather}
where the balancing condition \eqref{balanca} holds true.
Here the simultaneous choice of the integration contours as the real axis is valid under the constraints
$\operatorname{Im}(a_k)<0$ for all $k$ together with $\operatorname{Im}\big(a_k-\frac12 X\big)<1$, $k=1,2,3,4,$ and
$\operatorname{Im}\big(a_k+\frac12 X\big)<-1$, $k=5,6,7,8$.
In this relation we have two discrete parameters $\nu, \mu=0,\frac{1}{2}$.
If the integer~$L$ is even, then one has $\mu=\nu$. If~$L$ is an odd integer, then $\mu\neq \nu$.
This is completely similar to the~$W(E_7)$ group generating transformation for
the rarefied elliptic hypergeometric function derived in~\cite{spi:rare}
(see also~\cite{spi:rareYBE}). Analytical continuation
of the functions standing in~\eqref{complextrafoI} to other domains of parameters
can be reached by proper deformations of the integration contours. We note that the symmetry
transformation \eqref{complextrafoI} is a general complex analogue of the four term
Bailey transformation for non-terminating hypergeometric $_9F_8$-series.

\section{Transformation rule II}

The second type of identities follows from \eqref{vt1} after a group action composition,
\begin{equation}
V(t_1,\ldots,t_8;p,q)=\prod_{j, k=1}^4\Gamma(t_jt_{k+4};p,q)
V\Big(\tfrac{T^{1/2}}{t_1},\ldots,\tfrac{T^{1/2}}{t_4},\tfrac{U^{1/2}}{t_5},\ldots,\tfrac{U^{1/2}}{t_8};p,q\Big),
\label{vt2}\end{equation}
where $T=t_1t_2t_3t_4$ and $U=t_5t_6t_7t_8$.
The hyperbolic degeneration limit~\eqref{parlim2} for integrals described in the previous sections
reduces \eqref{vt2} to the following identity for $I_h(\underline{g})$ function
\begin{gather}\label{ide2}
I_h(\underline{g})=\prod_{j,k=1}^4\gamma^{(2)}(g_j+g_{k+4};\mathbb{\omega})
I_h(G-g_1,\ldots,G-g_4,Q-G-g_5,\ldots,Q-G-g_8),
\end{gather}
where $G:={1\over 2}\sum\limits_{j=1}^4 g_j$ and $Q=\omega_1+\omega_2$.

Let us apply a further degeneration limit to the complex hypergeometric integrals.
For $a_k$ and $N_k$ satisfying the balancing condition (\ref{balanca}), we denote
\begin{equation}
Y_1=\sum_{j=1}^4 a_j,\qquad L_1=\sum_{j=1}^4 N_j,\qquad Y_2=\sum_{j=5}^8 a_j,\qquad L_2=\sum_{j=5}^8 N_j,
\end{equation}
so that $Y_1+Y_2=-4{\rm i}$, $L_1+L_2=0$. Now the asymptotic relation \eqref{gam2lim2} and the
arguments given in previous sections reduce \eqref{ide2} to
\begin{gather}
\sum_{N\in {\mathbb Z}+\nu}\int_{-\infty}^{\infty}\big(y^2+N^2\big)
\prod_{k=1}^8{\bf \Gamma}(a_k\pm y,N_k\pm N){\rm d}y=
(-1)^{L_1}\prod_{j,k=1}^4{\bf \Gamma}(a_j+a_{k+4},N_j+N_{k+4})
\nonumber\\
\qquad{} \times
\sum_{N\in {\mathbb Z}+\mu}\int_{-\infty}^{\infty}\big(y^2+N^2\big)
\prod_{k=1}^4{\bf \Gamma}\big(\tfrac{1}{2}Y_1-a_k\pm y,\tfrac{1}{2}L_1-N_k\pm N\big)\nonumber
\\ \qquad{} {} \times
\prod_{k=5}^8{\bf \Gamma}\big(\tfrac{1}{2}Y_2-a_k\pm y, \tfrac{1}{2}L_2-N_k\pm N\big){\rm d}y.
\label{cide2}\end{gather}
Here we have $\mu=\nu$ for even integers $L_1$ (and, so, even $L_2$ as well), whereas $\mu\neq \nu$
for odd~$L_1$. The contours of integration can be taken as the real axis, provided imaginary parts of
the continuous parameters entering arguments of the complex gamma functions in~\eqref{cide2} are negative.

\section{Transformation rule III}

The third form of the symmetry transformation for the $V$-function follows from equating right-hand
side expressions in \eqref{vt1} and \eqref{vt2},
\begin{gather}
V(t_1,\ldots,t_8;p,q)=\prod_{1\leq j< k\leq 8}\Gamma(t_jt_{k};p,q)
V\left(\frac{\sqrt{pq}}{t_1},\ldots,\frac{\sqrt{pq}}{t_8};p,q\right).
\end{gather}
The hyperbolic degeneration limit (\ref{parlim2}) brings the following relation
for the $I_h(\underline{g})$ function
\begin{equation}\label{reflt}
I_h(\underline{g})=\prod_{1\leq j< k\leq 8}\gamma^{(2)}(g_j+g_k;\omega_1,\omega_2)
I_h\left(\underline{\lambda}\right), \qquad \lambda_j =\frac{\omega_1+\omega_2}{2}-g_j.
\end{equation}
Now we use the parametrisation \eqref{om1om22}, \eqref{zkom} and \eqref{gkom}, and
consider the $\delta\to 0^+$ limit \eqref{gam2lim2}. Then we have again
the balancing condition (\ref{balanca}) and the relation
\begin{equation}
{\omega_1+\omega_2\over 2}-g_k
={\rm i}\sqrt{\omega_1\omega_2}(-N_k+\delta(-a_k-{\rm i}))+O\big(\delta^2\big),\qquad k=1,\ldots,8.
\end{equation}
As a result of the same steps as in previous cases, we obtain the formula
\begin{gather}
\sum_{N\in {\mathbb Z}+\nu}\int_{-\infty}^{\infty}\big(y^2+N^2\big)
\prod_{k=1}^8{\bf \Gamma}(a_k\pm y,N_k\pm N){\rm d}y
=\prod_{1\leq j< k\leq 8}{\bf \Gamma}( a_j+a_k,N_j+N_k)
\nonumber\\ \qquad{} \times
\sum_{N\in {\mathbb Z}+\nu}\int_{-\infty}^{\infty}\big(y^2+N^2\big)
\prod_{k=1}^8{\bf \Gamma}(-{\rm i}-a_k\pm y,-N_k\pm N){\rm d}y,
\end{gather}
where $-1<\operatorname{Im}(a_k)<0$.

\section{Limiting case of the beta integral I}

The beta integral (\ref{hyper}) can serve as a source of many other calculable integrals
with a smaller number of the hyperbolic gamma functions in the kernel.
To derive them one should take to infinity some of the parameters $g_k$
in a smart way and use the asymptotic behaviour~(\ref{asy1}) and~(\ref{asy2}).
Here we will consider a couple of examples.

Let us set in (\ref{hyper})
\begin{equation}
g_j=f_j+{\rm i}\xi,\qquad g_{j+3}=h_j-{\rm i}\xi,\qquad j=1,2,3,\qquad \sum_{j=1}^3 (f_j+h_j)=Q,
\end{equation}
and also shift the integration variable $z\to z-{\rm i}\xi$.
Now we take the limit $\xi\to -\infty$ using the asymptotics (\ref{asy1}) and (\ref{asy2})
and obtain
\begin{equation}\label{hyper2}
\int_{-{\rm i}\infty}^{{\rm i}\infty}\prod_{j=1}^3\gamma^{(2)}(f_j+z;\mathbf{\omega})
\gamma^{(2)}(h_j-z;\mathbf{\omega}){dz\over {\rm i}\sqrt{\omega_1\omega_2}}
=\prod_{j, k =1}^3 \gamma^{(2)}(f_j+h_k;\mathbf{\omega}).
\end{equation}
Let us apply now the parametrisation \eqref{om1om22} and set in (\ref{hyper2})
\begin{equation}
f_j={\rm i}\sqrt{\omega_1\omega_2}(N_j+\delta s_j),
\qquad h_j={\rm i}\sqrt{\omega_1\omega_2}(M_j+\delta t_j),\qquad j=1,2,3,
\end{equation}
and $z={\rm i}\sqrt{\omega_1\omega_2}(N+\delta y)$, where $N, N_j, M_j \in {\mathbb Z}+\nu$, $\nu=0,\frac{1}{2}$. As a result, the balancing condition takes the form
\[
\sum_{j=1}^3( N_j+M_j)=0,\qquad \sum_{j=1}^3( s_j+t_j)=-2{\rm i} .
\]
In the limit $\delta\to 0^+$, using the arguments of previous sections and formula (\ref{gam2lim2}),
one can see that relation (\ref{hyper2}) reduces to
\begin{gather}
\frac{1}{4\pi} \sum_{N\in {\mathbb Z}+\nu}
\int_{-\infty}^{\infty}\prod_{j=1}^3{\bf \Gamma}(s_j+y,N+N_j)
{\bf \Gamma}(t_j-y,N-M_j){\rm d}y\nonumber\\
 \qquad{} =\prod_{j,k=1}^3{\bf \Gamma}(s_j+t_k,N_j+M_k). \label{limtrafoI}
 \end{gather}
For $\nu=0$ this identity was obtained earlier in \cite{BMS} for special values
of $N_j$, $M_j$ and in \cite{DMV2017} it was derived for general $N_j$, $M_j$. As shown in the
latter paper, for $\nu=0$ this is nothing else than the Mellin--Barnes representation of the
general complex star-triangle relation~\eqref{STR}.
The case $\nu=1/2$ of formula \eqref{limtrafoI} reduces to the case $\nu=0$ by the shifts $N_j\to N_j+\nu$, $M_k\to M_k-\nu$.

\section{Limiting case of the beta integral II}

Consider now the following limit in the identity (\ref{hyper})
\[
g_5\to {\rm i}\infty,\qquad g_6=Q-g_5-\sum_{j=1}^4g_j\to -{\rm i}\infty.
\]
Using the asymptotics (\ref{asy1}) and (\ref{asy2}), we come to the hyperbolic
analogue of the Askey--Wilson $q$-beta integral established by Ruijsenaars \cite{Ruij}
\begin{equation}\label{hyper4}
 \int_{-{\rm i}\infty}^{{\rm i}\infty}{\prod\limits_{k=1}^4\gamma^{(2)}(g_k\pm z;\mathbf{\omega})\over \gamma^{(2)}(\pm 2z;\mathbf{\omega})}{{\rm d}z\over 2{\rm i}\sqrt{\omega_1\omega_2}}=
{\prod\limits_{1\leq j < k \leq 4}
\gamma^{(2)}(g_j+g_k;\mathbf{\omega})\over \gamma^{(2)}\Big(\sum\limits_{k=1}^ 4 g_k;\mathbf{\omega}\Big)}.
\end{equation}
Setting
\begin{equation}
{g_k\over {\rm i}\sqrt{\omega_1\omega_2}}=N_k+\delta a_k,\qquad k=1,2,3,4, \qquad
\frac{z}{{\rm i}\sqrt{\omega_1\omega_2}}=N+\delta y,
\end{equation}
and using the same limit $\delta\to 0^+$ as before (\ref{gam2lim2}), one can see that (\ref{hyper4})
reduces to
\begin{gather} \nonumber
{1\over 8\pi}\sum_{N\in {\mathbb Z}+\nu}\int_{-\infty}^{\infty}\big(y^2+N^2\big)\prod_{k=1}^4{\bf \Gamma}(a_k\pm y,N_k\pm N)
\\ \qquad{}
=(-1)^{2\nu}{\prod\limits_{1\leq j < k \leq 4}
{\bf \Gamma}(a_j+a_k,N_j+N_k)\over {\bf \Gamma}\Big(\sum\limits_{k=1}^ 4 a_k,\sum\limits_{k=1}^ 4 N_k\Big)}.
\label{hyper44}\end{gather}
This is a general complex analogue of the de Branges--Wilson integral \cite{aar}.
For $\nu=0$ (i.e., for integer values of $N$) and all $N_k=0$ this relation
was obtained in \cite{Neretin2018} (with the contradiction that instead of our $1/8\pi$
factor on the left-hand side there stands $1/4\pi^2$).\footnote{Exactly the same relation \eqref{hyper44} was obtained also independently in \cite{DM2019} as formula (2.3b) for $N=2$.}

\section{Limiting case of the transformation rules I}

Obviously, the same limiting transitions can be performed also for the transformation rules~(\ref{ide1}), (\ref{ide2}), and~(\ref{reflt}). In this way we will obtain a number of new relations between integrals with a~smaller number of hyperbolic gamma functions. Let us apply this procedure to the last rule~(\ref{reflt}).
It is straightforward to do the same with relations~(\ref{ide1}) and~(\ref{ide2}), but we skip them
for brevity. So, consider the limit
\[
g_7\to {\rm i}\infty,\qquad g_8=2Q-\sum_{j=1}^6g_j-g_7\to -{\rm i}\infty
\]
in (\ref{reflt}). As a result we obtain
\begin{gather} \nonumber
\int_{-{\rm i}\infty}^{{\rm i}\infty}{\prod\limits_{j=1}^6\gamma^{(2)}(g_j\pm z;\mathbf{\omega})\over
\gamma^{(2)}(\pm 2z;\mathbf{\omega})}{\rm d}z
\\ \qquad{} = {1\over \gamma^{(2)}(G-Q;\mathbf{\omega})}
\prod_{1\leq j< k \leq 6}\gamma^{(2)}(g_j+g_k;\mathbf{\omega})
\int_{-{\rm i}\infty}^{{\rm i}\infty}{\prod\limits_{j=1}^6\gamma^{(2)}(\tilde{g_j}\pm z;\mathbf{\omega})\over
\gamma^{(2)}(\pm 2z;\mathbf{\omega})}{\rm d}z,
\label{gmro} \end{gather}
where
\begin{equation}
\tilde{g_j}=\tfrac{1}{2}Q-g_j, \qquad j=1,\ldots,6,\qquad
G=\sum_{j=1}^6g_j, \qquad Q=\omega_1+\omega_2.
\end{equation}
Taking the parametrisation of variables \eqref{om1om2}, \eqref{zkom}, and \eqref{gkom},
we also have
\[
\frac{G-Q}{{\rm i}\sqrt{\omega_1\omega_2}}
=\sum_{k=1}^6 N_k+\delta\left(\sum_{k=1}^6a_k+2{\rm i}\right)+O\big(\delta^2\big).
\]
Now, in the limit $\delta\to 0^+$ we use formula (\ref{gam2lim2}), and then relation (\ref{gmro})
reduces to\footnote{This formula corresponds to the choice $n=m=1$ in formula (6.7) in~\cite{DM2019}.}
\begin{gather} \nonumber
\sum_{N\in {\mathbb Z}+\nu} \int_{-\infty}^{\infty}\big(y^2+N^2\big)\prod_{j=1}^6{\bf \Gamma}(a_j\pm y,N_j\pm N){\rm d}y
\\ \qquad{}
=(-1)^{2\nu}{\prod\limits_{1\leq j< k \leq 6}{\bf \Gamma}(a_j+a_k, N_j+N_k)
\over \Gamma\Big(\sum\limits_{j=1}^6 a_j+2{\rm i},\sum\limits_{j=1}^6 N_j\Big)}\nonumber
\\ \qquad\quad{} \times
\sum_{N\in {\mathbb Z}+\nu} \int_{-\infty}^{\infty}\big(y^2+N^2\big)
\prod_{j=1}^6{\bf \Gamma}(-{\rm i}-a_j\pm y,-N_j\pm N){\rm d}y,
\label{degtrafo1}\end{gather}
where we assume that the contours of integration are either real axes
for $-1<\textup{Im}(a_k)<0$ or their proper deformations allowing analytical
continuations of the functions on both sides in the variables $a_k$.

\section{Limiting case of transformation rules II}

For deriving another symmetry transformation, we replace in (\ref{reflt})
\[
g_j \to g_j+{\rm i}\xi, \qquad g_{j+4}=f_j-{\rm i}\xi, \qquad j=1,\ldots, 4,
\qquad z\to z-{\rm i}\xi.
\]
The balancing condition takes the form $\sum\limits_{j=1}^4 (f_j+g_j)=2Q$.
After taking the limit $\xi\to -\infty$, we come to the identity
\begin{gather}\nonumber
\int_{-{\rm i}\infty}^{{\rm i}\infty}\prod_{j=1}^4\gamma^{(2)}(g_j+z;\mathbf{\omega})
\gamma^{(2)}(f_j-z;\mathbf{\omega}) {\rm d}z
=\prod_{j,k=1}^4 \gamma^{(2)}(g_j+f_k;\mathbf{\omega})
\\ \qquad{} \times
\int_{-{\rm i}\infty}^{{\rm i}\infty}\prod_{j=1}^4\gamma^{(2)}\big(\tfrac{1}{2}Q-f_j+z;\mathbf{\omega}\big)
\gamma^{(2)}(\tfrac{1}{2}Q-g_j-z;\mathbf{\omega}) {\rm d}z.
\label{infy} \end{gather}
Now we take the parametrisation \eqref{om1om2}, \eqref{zkom} jointly with
\begin{equation}
\frac{g_j}{{\rm i}\sqrt{\omega_1\omega_2}}=N_j+\delta s_j,\qquad \frac{f_j}{{\rm i}\sqrt{\omega_1\omega_2}}=M_j+\delta t_j,\qquad j=1,2,3,4.
\label{newpar}\end{equation}
The balancing condition takes the form
\begin{equation}
\sum_{j=1}^4( N_j+M_j)=0,\qquad \sum_{j=1}^4(s_j+t_j)=-4{\rm i}.
\end{equation}
In the limit $\delta\to 0^+$, in the same way as in many cases before, equation
(\ref{infy}) yields the following identity
\begin{gather}
\sum_{N\in {\mathbb Z}+\nu}\int_{-\infty}^{\infty}\prod_{k=1}^4{\bf \Gamma}(s_k+y,N_k+N)
{\bf \Gamma}(t_k-y,M_k-N) {\rm d}y
\nonumber\\ 
\qquad{}=(-1)^{\sum\limits_{k=1}^4N_k}\prod_{j,k=1}^4 {\bf \Gamma}(s_j+t_k,N_j+M_k)
\nonumber\\ \qquad\quad{} \times
\sum_{N\in {\mathbb Z}+\nu}\int_{-\infty}^{\infty}\prod_{k=1}^4{\bf \Gamma}(-{\rm i}-t_k+y,N-M_k)
{\bf \Gamma}(-{\rm i}-s_k-y,-N-N_k) {\rm d}y.
 \end{gather}

After resolving the balancing condition in favor of $f_4$,
\[
f_4=2Q-g_4-\sum_{k=1}^3(f_k+g_k),
\]
and taking the limit $g_4\to {\rm i}\infty$ in (\ref{infy}), we obtain
\begin{gather}
\int_{-{\rm i}\infty}^{{\rm i}\infty}{\rm e}^{{\pi{\rm i} z\over \omega_1\omega_2}\big(Q-\sum\limits_{k=1}^3(f_k+g_k)\big)}\prod_{k=1}^3\gamma^{(2)}(g_k+z;\mathbf{\omega})
\gamma^{(2)}(f_k-z;\mathbf{\omega}){\rm d}z
\nonumber\\ \qquad {}
={\rm e}^{{\pi {\rm i}\over 2\omega_1\omega_2}\big(Q\sum\limits_{k=1}^3(f_k-g_k)+2\sum\limits_{1\leq j< k\leq 3}(g_jg_k-f_jf_k)\big)} {\prod\limits_{j,k=1}^3
\gamma^{(2)}(g_j+f_k;\mathbf{\omega})\over \gamma^{(2)}\Big(\sum\limits_{k=1}^3(g_k+f_k)-Q;\mathbf{\omega}\Big)}
\\ \qquad\quad {} \times
\int_{-{\rm i}\infty}^{{\rm i}\infty}{\rm e}^{{\pi {\rm i}z\over \omega_1\omega_2}\big({-}2Q+\sum\limits_{k=1}^3(g_k+f_k)\big)}\prod_{k=1}^3\gamma^{(2)}\big(\tfrac{1}{2}Q-f_k+z;\mathbf{\omega}\big)
\gamma^{(2)}\big(\tfrac{1}{2}Q-g_k-z;\mathbf{\omega}\big) {\rm d}z.\nonumber
\end{gather}
Taking the same parametrization \eqref{om1om2}, \eqref{zkom} and \eqref{newpar},
the limit $\delta\to 0^+$ yields the identity\footnote{This identity was also obtained in \cite{DM2019}, see there formula (6.6) for $n=m=1$.}
\begin{gather}
\sum_{N\in \mathbb{Z}+\nu} (-1)^{N-\nu} \int_{-\infty}^{\infty}\prod_{k=1}^3{\bf \Gamma}(y+s_k,N+N_k)
{\bf \Gamma}(-y+t_k,M_k-N)
{\rm d}y\nonumber\\ \qquad{}
={(-1)^{\sum\limits_{k=1}^3(N_k+M_k)}\prod\limits_{k,j=1}^3
{\bf \Gamma}(s_k+t_j,N_k+M_j)\over {\bf \Gamma}\Big(\sum\limits_{k=1}^3(s_k+t_k)+2{\rm i},\sum\limits_{k=1}^3(N_k+M_k)\Big)}
\nonumber\\ \qquad\quad{}
\times\sum_{N\in \mathbb{Z}+\nu}\int_{-\infty}^{\infty}(-1)^{N-\nu}
\prod_{k=1}^3{\bf \Gamma}(y-{\rm i}-t_k, N-M_k) {\bf \Gamma}(-y-{\rm i}-s_k,-N-N_k) {\rm d}y.
\end{gather}
It is easy to see that for $\nu=1/2$ this formula is identical with the $\nu=0$ case
after replacing~$N_k$ by~$N_k-1$, i.e., the parameter $\nu$ becomes redundant.
As usual, the contours of integration separate sequences of poles of the integrands going upwards
from the ones falling down.

\section[A new degeneration of the hyperbolic gamma function ($\omega_1=\omega_2$)]{A new degeneration of the hyperbolic gamma function\\ ($\boldsymbol{\omega_1=\omega_2}$)}\label{newlimit}

Now we consider the limit $\omega_2\to\omega_1$, or $b\to 1$
corresponding to the central charge $c=25$ (the case $b\to -1$ is equivalent to it and we skip it).
In this case, for generic values of~$u$ the hyperbolic gamma function \eqref{int_rep} (or \eqref{HGF})
remains a well defined meromorphic function of $u$. However, for special choices of $u$ we have
a divergence which we describe below. Namely, we take small $\delta>0$ and set
\begin{equation}
b=\sqrt{\omega_1\over \omega_2}=1+{\rm i}\delta.
\end{equation}
As a consequence, $Q=\omega_1+\omega_2=2\sqrt{\omega_1\omega_2}+O\big(\delta^2\big)$ and
\[
\sqrt{\omega_2\over \omega_1}=1-{\rm i}\delta+O\big(\delta^2\big),\qquad
{\omega_1\over \omega_2}=1+2{\rm i}\delta-\delta^2,\qquad {\omega_2\over \omega_1}=1-2{\rm i}\delta+O\big(\delta^2\big).
\]
Also we parametrise the argument $u$ in \eqref{hypergamma} as
\begin{equation}
u=\sqrt{\omega_1\omega_2}(n+y\delta), \qquad n\in {\mathbb Z}, \qquad y\in {\mathbb C},
\end{equation}
and consider the limit $\delta\to 0^+$.
For infinite products entering \eqref{hypergamma} we have
\[
\big({\rm e}^{2\pi {\rm i} \frac{u}{\omega_2}};{\rm e}^{2\pi {\rm i} {\omega_1\over \omega_2}}\big)_{\infty}
=\big(q^{\frac{n-{\rm i}y}{2}+O(\log q)};q\big)_\infty
=\frac{(q;q)_\infty(1-q)^{1+\frac{-n+{\rm i}y}{2}+O(\log q)}}
{\Gamma_q\big(\frac{n-{\rm i}y}{2}+O(\log q)\big)},
\]
where $q={\rm e}^{-4\pi\delta(1+{\rm i}\delta/2)}$. Analogously,
\[
\big({\rm e}^{2\pi {\rm i}\frac{u}{\omega_1}}{\rm e}^{-2\pi {\rm i} {\omega_2\over \omega_1}};{\rm e}^{-2\pi {\rm i} {\omega_2\over \omega_1}}\big)_{\infty}
= \big(\tilde q^{1-\frac{n+{\rm i}y}{2}+O(\log \tilde q)};\tilde q\big)_\infty
=\frac{(\tilde q;\tilde q)_\infty(1-\tilde q)^{\frac{n+{\rm i}y}{2}+O(\log \tilde q)}}
{\Gamma_{\tilde q}\big(1-\frac{n+{\rm i}y}{2}+O(\log \tilde q)\big)},
\]
where
\[
\tilde q={\rm e}^{-4\pi\delta\frac{1+{\rm i}\delta/2}{(1+{\rm i}\delta)^2}}.
\]
As a result we obtain,
\[
\gamma(u;\omega_1,\omega_2)=\frac{\Gamma_q\big( \frac{n-{\rm i}y}{2}+O(\log q) \big)}
{\Gamma_{\tilde q}\big( 1-\frac{n+{\rm i}y}{2}+O(\log \tilde q)\big)}
\frac{(\tilde q;\tilde q)_\infty}{(q;q)_\infty}
\frac{(1-\tilde q)^{\frac{n+{\rm i}y}{2}+O(\log \tilde q)}}
{(1-q)^{1+\frac{-n+{\rm i}y}{2}+O(\log q)}}.
\]

Using the modular transformation rule for the Dedekind eta-function, we find
\begin{equation}
\frac{(\tilde q;\tilde q)_\infty}{(q;q)_\infty}
={\rm e}^{{\pi{\rm i} \over 12}\big({\omega_2\over \omega_1}+{\omega_1\over \omega_2}\big)}
\left(-{\rm i} {\omega_1\over \omega_2}\right)^{1\over 2}
\underset{\delta\to 0^+}{=}{\rm e}^{ -\frac{\pi{\rm i}}{12}}.
\end{equation}
Finally, applying the strong limit $\Gamma_q(x+O(\log q))\to \Gamma(x)$ for $q\to 1$,
which is uniform on compacta~\cite{Tom,rai:limits},
and combining all factors together, we come to the leading asymptotics
\begin{equation}
\gamma(u;\omega_1,\omega_2)\underset{\delta\to 0^+}{=} {\rm e}^{-\frac{\pi {\rm i}}{12}}
(4\pi\delta)^{n-1} \left(1-\frac{n+{\rm i}y}{2}\right)_{n-1}, \qquad (a)_n:=\frac{\Gamma(a+n)}{\Gamma(a)},
\end{equation}
where $(a)_0=1$ and
\[
(a)_n=
\begin{cases}
 a(a+1)\cdots(a+n-1), & \text{for} \ n>0,\quad \\
\dfrac{1}{(a-1)(a-2)\cdots(a+n)}, &\text{for} \ n<0,
\end{cases}
\]
is the standard Pochhammer symbol.
For function \eqref{HGF} this yields
\[
\gamma^{(2)}(u;\omega_1,\omega_2)\underset{\delta\to 0^+}{=} {\rm e}^{-\frac{\pi {\rm i}}{2}(n-1)^2}
(4\pi\delta)^{n-1}\left(1-\frac{n+{\rm i}y}{2}\right)_{n-1}.
\]
Since here $n$ is an arbitrary integer, we can shift $n\to n+1$ in this formula, which
noticeably simplifies its form
\begin{equation}
\gamma^{(2)}\big(\sqrt{\omega_1\omega_2}(n+1+y\delta);\mathbf{\omega}\big)
\underset{\delta\to 0^+}{=} {\rm e}^{-\frac{\pi {\rm i}}{2}n^2}
(4\pi\delta)^n\left(\frac{1-n-{\rm i}y}{2}\right)_n,
\qquad \sqrt{\omega_1\over \omega_2}=1+{\rm i}\delta,
\end{equation}
where $n\in {\mathbb Z}$, $y\in {\mathbb C}$.
This is a new degeneration limit for the hyperbolic gamma function or the Faddeev modular quantum dilogarithm.
In this case the function $\gamma^{(2)}(u;\mathbf{\omega})$
starts either to vanish or to blow up around a special discrete set of points of the argument $u$
passing through the whole complex plane along a particular line. This is partially similar
to the picture taking place for $\omega_1+\omega_2\to 0$, but the direction of
the corresponding line on the complex plane is different.
We postpone consideration of the consequences of such a degeneration for hyperbolic beta integrals
and its possible applications to a later work.

\section{Conclusion}

In the present paper we have performed a rigorous and complete analysis of the degeneration
of hyperbolic integrals to the complex hypergeometric functions in the Mellin--Barnes
representation, which was noticed for the first time in \cite{BMS}. The limit
$\omega_1+\omega_2\to 0$ corresponds to $b\to \pm {\rm i}$ in the context of $2d$
Liouville quantum field theory (which is related to the $p\to\infty$ limit of minimal
models leading to the central charge value $c=1$ \cite{RW}).
 Additionally, we have discovered a~new nontrivial
degeneration of the hyperbolic gamma function, or Faddeev's modular quantum dilogarithm,
in the limit $\omega_1\to\omega_2$, which
corresponds to $b\to\pm1$ for the Liouville theory leading to the central charge $c=25$.
It would be interesting to investigate applications of our formulas in the context
of fusion matrices of the corresponding two-dimensional quantum field theories.

Another interesting application should emerge within the theory of Painlev\'e transcendents.
Namely, as follows from the analysis of $c=1$ conformal blocks in \cite{ILT},
the degeneration of the function standing in the left-hand side of the identity
\eqref{complextrafoI} to the one in \eqref{degtrafo1} should define the fusion matrix
of interest with a direct relation
to a particular tau-function of the Painlev\'e-VI function. Similar question can be raised
for the $c=25$ conformal blocks and their manifestations for the Painlev\'e equations.

The third application in the quantum field theory is expected to emerge in the topological
field theory. Namely, one can extend considerations of \cite{KLV} and realize the
corresponding Pachner moves using the unusual integral identities derived by us.

From the point of view of
representation theory of the complex group ${\rm SL}(2,{\mathbb C})$, it is necessary to clarity the
group-theoretical meaning of the parameter value $\nu=1/2$, showing very interesting
phenomena \cite{spi:rare}. In this context, one of the unsolved problems is the
inversion of the Mellin--Barnes form of our functions -- infinite bilateral sums of integrals
standing in identities~\eqref{keybeta},~\eqref{complextrafoI}, etc.~-- to the integrals
over complex planes of the type \eqref{CB} and \eqref{STR}. For $\nu=0$ some examples of such
conversions are given in \cite{DMV2017,DS2017,Ismag2}. However, for $\nu=1/2$ it is not clear
how it can be performed yet.

There are also applications to integrable systems. Namely, since hyperbolic beta integrals
serve as measures of the orthogonality relations for wave functions of the Ruijsenaars type many body
systems \cite{Ruij}, one can consider what happens with them in the taken limits.
In particular, this requires construction of proper generalizations of the hypergeometric
equation to a finite-difference equation for the function standing in the left-hand side
of equality \eqref{complextrafoI}. As another manifestation in integrable systems, it is
straightforward to build the corresponding solution of the Yang--Baxter
equation just by appropriate degeneration of the results of \cite{CS}
(or, in a more general setting, the results of \cite{spi:rareYBE}).

Obviously, the degenerations similar to the one we considered in the present paper for
the most general univariate hyperbolic beta integral
and symmetry transformations for the corresponding analogue
of the Euler--Gauss hypergeometric function can be applied to the rarefied hyperbolic
integrals associated with the general lens space \cite{SS23, SS24}.
In particular, the first task would be to consider the most general way of
approaching the unit circle simultaneously by $q$ and $\tilde q$ in the rarefied $q$-beta
integral \eqref{integral}.

Finally, we have presented the degeneration hierarchy
only for the simplest identities for elliptic hypergeometric integrals \cite{spi:umn,spi:essays}
and one can extend our approach to all relations between multidimensional integrals
of such type established to the present moment.

\subsection*{Acknowledgements}
This paper is based on the talk given by V.S.\ at the conference ``Elliptic Integrable Systems,
Special Functions and Quantum Field Theory'', June 16--20, 2019, Nordita, Stockholm.
The key results of this work were obtained within the research program of project no.~19-11-00131
supported by the Russian Science Foundation.
We thank T.H.~Koornwinder and E.M.~Rains for explanations on the
uniformness of the limit for $q$-gamma function \eqref{gamlim}
following from their works~\cite{Tom} and~\cite{rai:limits}.

\pdfbookmark[1]{References}{ref}
\LastPageEnding

\end{document}